\documentclass[12pt]{article}

\usepackage{axodraw}
\usepackage{epsfig}

\makeatletter

\long\def\@caption#1[#2]#3{%
  \par
  \addcontentsline{\csname ext@#1\endcsname}{#1}%
    {\protect\numberline{\csname the#1\endcsname}{\ignorespaces #2}}%
  \begingroup
    \@parboxrestore
    \if@minipage
      \@setminipage
    \fi
    \normalsize
    \@makecaption{\csname fnum@#1\endcsname}{\it \ignorespaces #3}\par
  \endgroup}

\def\paragraph{\@startsection{paragraph}{4}{\z@}{+2.00ex plus
 +1ex minus +.2ex}{1.5ex plus .2ex}{\it\normalsize}}

\def\section{\@startsection {section}{1}{\z@}{+3.0ex plus +1ex minus
  +.2ex}{2.3ex plus .2ex}{\normalsize\bf\boldmath}}
\def\subsection{\@startsection{subsection}{2}{\z@}{+2.5ex plus +1ex
minus +.2ex}{1.5ex plus .2ex}{\normalsize\bf\boldmath}}
\def\subsubsection{\@startsection{subsubsection}{3}{\z@}{+3.25ex plus
 +1ex minus +.2ex}{1.5ex plus .2ex}{\normalsize\bf\boldmath}}

\expandafter\ifx\csname mathrm\endcsname\relax\def\mathrm#1{{\rm #1}}\fi

\newcounter{saveeqn}

\@addtoreset{equation}{section}

\newcount\@tempcntc
\def\@citex[#1]#2{\if@filesw\immediate\write\@auxout{\string\citation{#2}}\fi
  \@tempcnta\z@\@tempcntb\m@ne\def\@citea{}\@cite{\@for\@citeb:=#2\do
    {\@ifundefined
       {b@\@citeb}{\@citeo\@tempcntb\m@ne\@citea
        \def\@citea{,\penalty\@m\ }{\bf ?}\@warning
       {Citation `\@citeb' on page \thepage \space undefined}}%
    {\setbox\z@\hbox{\global\@tempcntc0\csname
b@\@citeb\endcsname\relax}%
     \ifnum\@tempcntc=\z@ \@citeo\@tempcntb\m@ne
       \@citea\def\@citea{,\penalty\@m}
       \hbox{\csname b@\@citeb\endcsname}%
     \else
      \advance\@tempcntb\@ne
      \ifnum\@tempcntb=\@tempcntc
      \else\advance\@tempcntb\m@ne\@citeo
      \@tempcnta\@tempcntc\@tempcntb\@tempcntc\fi\fi}}\@citeo}{#1}}

\def\@citeo{\ifnum\@tempcnta>\@tempcntb\else\@citea
  \def\@citea{,\penalty\@m}%
  \ifnum\@tempcnta=\@tempcntb\the\@tempcnta\else
   {\advance\@tempcnta\@ne\ifnum\@tempcnta=\@tempcntb \else
\def\@citea{--}\fi
    \advance\@tempcnta\m@ne\the\@tempcnta\@citea\the\@tempcntb}\fi\fi}

\newcommand{\lsim}
{\mathrel{\raisebox{-.3em}{$\stackrel{\displaystyle <}{\sim}$}}}

\def\asymp#1%
{\mathrel{\raisebox{-.4em}{$\widetilde{\scriptstyle #1}$}}}

\def\Nequal#1%
{\mathrel{\raisebox{-.5em}{$\stackrel{=}{\scriptstyle\rm#1}$}}}
\newcommand{\dsl}[1]{\not \hspace{-0.7mm}#1}
\def\dsl{\mathpalette\make@slash}
\def\make@slash#1#2{\setbox\z@\hbox{$#1#2$}%
  \hbox to 0pt{\hss$#1/$\hss\kern-\wd0}\box0}

\def\beq{\begin{equation}}
\def\eeq{\end{equation}}
\def\beqar{\begin{eqnarray}}
\def\eeqar{\end{eqnarray}}
\def\barr#1{\begin{array}{#1}}
\def\earr{\end{array}}
\def\bfi{\begin{figure}}
\def\efi{\end{figure}}
\def\btab{\begin{table}}
\def\etab{\end{table}}
\def\bce{\begin{center}}
\def\ece{\end{center}}

\def\text{\textstyle}

\def\la{\lambda}

\def\refeq#1{\mbox{(\ref{#1})}}

\def\reffi#1{\mbox{Fig.~\ref{#1}}}
\def\reffis#1{\mbox{Figs.~\ref{#1}}}

\def\citere#1{\mbox{Ref.~\cite{#1}}}
\def\citeres#1{\mbox{Refs.~\cite{#1}}}

\newcommand{\GeV}{\unskip\,\mathrm{GeV}}

\newcommand{\rd}{{\mathrm{d}}}

\def\mathswitchr#1{\relax\ifmmode{\mathrm{#1}}\else$\mathrm{#1}$\fi}
\def\mathswitch#1{\relax\ifmmode#1\else$#1$\fi}

\newcommand{\PH}{\mathswitch {H}}

\newcommand{\Pt}{\mathswitch {t}}

\newcommand{\MH}{\mathswitch {M_\PH}}

\def\ie{i.e.\ }

\newcommand{\LO}{\mathrm{LO}}
\newcommand{\NLO}{\mathrm{NLO}}

\renewcommand{\O}{{\cal O}}

\def\draftdate{\relax}
\def\mda{\relax}
\def\mua{\relax}
\def\mla{\relax}
\def\draft{
\def\thtystars{******************************}
\def\sixtystars{\thtystars\thtystars}
\typeout{}
\typeout{\sixtystars**}
\typeout{* Draft mode!
         For final version remove \protect\draft\space in source file *}
\typeout{\sixtystars**}
\typeout{}
\def\draftdate{\today}
\def\mua{\marginpar[\boldmath\hfil$\uparrow$]%
                   {\boldmath$\uparrow$\hfil}%
                    \typeout{marginpar: $\uparrow$}\ignorespaces}
\def\mda{\marginpar[\boldmath\hfil$\downarrow$]%
                   {\boldmath$\downarrow$\hfil}%
                    \typeout{marginpar: $\downarrow$}\ignorespaces}
\def\mla{\marginpar[\boldmath\hfil$\rightarrow$]%
                   {\boldmath$\leftarrow $\hfil}%
                    \typeout{marginpar: $\leftrightarrow$}\ignorespaces}
\def\Mua{\marginpar[\boldmath\hfil$\Uparrow$]%
                   {\boldmath$\Uparrow$\hfil}%
                    \typeout{marginpar: $\uparrow$}\ignorespaces}
\def\Mda{\marginpar[\boldmath\hfil$\Downarrow$]%
                   {\boldmath$\Downarrow$\hfil}%
                    \typeout{marginpar: $\downarrow$}\ignorespaces}
\def\Mla{\marginpar[\boldmath\hfil$\Rightarrow$]%
                   {\boldmath$\Leftarrow $\hfil}%
                    \typeout{marginpar: $\leftrightarrow$}\ignorespaces}
\overfullrule 5pt
\oddsidemargin -15mm
\marginparwidth 29mm
}

\def\stars{\strut\leaders\hbox{*}\hfill\strut}
\def\starline{\hfil\strut\hfil\hbox to \textwidth {\stars}\hfil}

\begin{document}
\thispagestyle{empty}
\def\thefootnote{\fnsymbol{footnote}}
\setcounter{footnote}{1}
\null
\draftdate\hfill MPI-PhT/2002-62 \\
\strut\hfill hep-ph/0210380

\vfill
\begin{center}

{\Large \bf  Associated Higgs production with heavy quark pairs 
\\[.5em]
at hadron colliders%
\footnote{To appear in the proceedings of 
{\it The 10th International Conference on Supersymmetry
and Unification of Fundamental Interactions, SUSY02},
June 2002, DESY Hamburg}
}

\vspace*{1cm}

{\sc Stefan Dittmaier}

\vspace*{.5cm}

{\normalsize \it
Max-Planck-Institut f\"ur Physik 
(Werner-Heisenberg-Institut) \\
F\"ohringer Ring 6, D-80805 M\"unchen, Germany}
\par
\end{center}
\vskip 2cm
\begin{center}
\bf Abstract
\end{center} 
The reactions $p\bar p/pp\to t\bar tH+X$ represent important channels
in the search for the Standard Model Higgs boson at the Tevatron and the LHC.
In the leading perturbative order the cross sections suffer from severe
(renormalization and factorization) scale uncertainties. 
The next-to-leading order QCD corrections stabilize the cross sections
considerably. The calculation of these corrections is briefly reviewed,
and a few numerical results are discussed.
\par
\vskip 1cm
\vfill
\noindent
October 2002
\vskip 1cm
\null
\setcounter{page}{0}

\clearpage

\vspace*{1cm}
\begin{center}

{\Large \bf  Associated Higgs production with heavy quark pairs 
\\[.5em]
at hadron colliders}

\vspace*{1cm}

{\sc Stefan Dittmaier}

\vspace*{.5cm}

{\normalsize \it
Max-Planck-Institut f\"ur Physik 
(Werner-Heisenberg-Institut) \\
F\"ohringer Ring 6, D-80805 M\"unchen, Germany}
\par
\end{center}
\vskip 1cm
\begin{center}
\bf Abstract
\end{center} 
{\it
The reactions $p\bar p/pp\to t\bar tH+X$ represent important channels
in the search for the Standard Model Higgs boson at the Tevatron and the LHC.
In the leading perturbative order the cross sections suffer from severe
(renormalization and factorization) scale uncertainties. 
The next-to-leading order QCD corrections stabilize the cross sections
considerably. The calculation of these corrections is briefly reviewed,
and a few numerical results are discussed.
}
\par
\vskip 1cm

\section{Introduction}
\label{se:intro}

One of the most pressing open questions in high-energy physics
concerns the mechanism of electroweak symmetry breaking. 
In the electroweak Standard Model (SM) and its extensions,
the symmetry breaking is realized spontaneously 
by the Higgs mechanism which predicts the existence of
Higgs bosons.
Thus, the search for Higgs bosons is among the first incentives
in present high-energy experiments.
For the SM Higgs boson, a lower experimental mass limit of
$\MH>114.4\GeV$ (95\% C.L.) has been set by the LEP experiments 
\cite{MHlow}.
Moreover, an upper bound of 
$\MH < 196$ GeV (95\% C.L.) \cite{Abbaneo:2001ix}
results from a fit of the SM parameters in the predictions for
electroweak precision data.

In the near future, the search for Higgs bosons will continue at
the $p\bar p$ collider Tevatron \cite{Carena:2000yx} and later at the
$pp$ collider LHC \cite{atlas_cms_tdrs}.
Among the various discovery channels for Higgs bosons in the intermediate 
mass range, Higgs-boson radiation off top quarks,
$p\bar p/pp\to\Pt\bar\Pt\PH+X$, plays an important r\^ole.
Although the expected rate is low at the Tevatron, a sample of a few
but very clean events could be observed for $\MH\lsim 140\GeV$
\cite{Goldstein:2000bp}. At the LHC, 
$\Pt\bar\Pt\PH$ production is an important search
channel for $\MH\lsim 125$ GeV \cite{Drollinger:2001ym}.
Moreover, analyzing the $\Pt\bar\Pt\PH$ production rate at the LHC can
provide valuable information on the top--Higgs Yukawa coupling 
\cite{Maltoni:2002jr}. 

Pure leading-order (LO) predictions for the 
$p\bar p/pp\to\Pt\bar\Pt\PH+X$ cross sections, which have been
available in the literature \cite{Kunszt:1984ri} for a long time, 
are notoriously imprecise, since they suffer from
considerable uncertainties owing to the strong dependence on the
renormalization and factorization scales. This uncertainty can
only be reduced by including higher-order QCD corrections.
More recently the complete calculation of the next-to-leading order (NLO)
QCD corrections to the total cross section at the Tevatron as well as the LHC
was presented in \citere{Beenakker:2001rj}. These results are in
agreement with the parallel calculation of \citere{Reina:2001sf}
where, however, only the $q\bar q$ annihilation subprocess, which dominates at 
the Tevatron, has been taken into account.
As expected, in the NLO predictions of the cross sections the
scale dependence is reduced significantly.

In this brief article the salient features of the calculation 
\cite{Beenakker:2001rj,pptth_long} of the
full NLO QCD corrections are summarized.
In the next section, the main obstacles in the actual calculation
and the basic ideas of their solution are explained; a detailed
description with explicit results is contained in \citere{pptth_long}.
Finally, a few examples for phenomenological results are discussed, including 
the scale dependence of the total cross sections at the Tevatron and 
the LHC as well as the transverse-momentum and rapidity distributions of the
Higgs boson at the Tevatron; more detailed results can be found in
\citeres{Beenakker:2001rj,pptth_long}.

\section{Features of the NLO calculation}

\paragraph{General remarks}

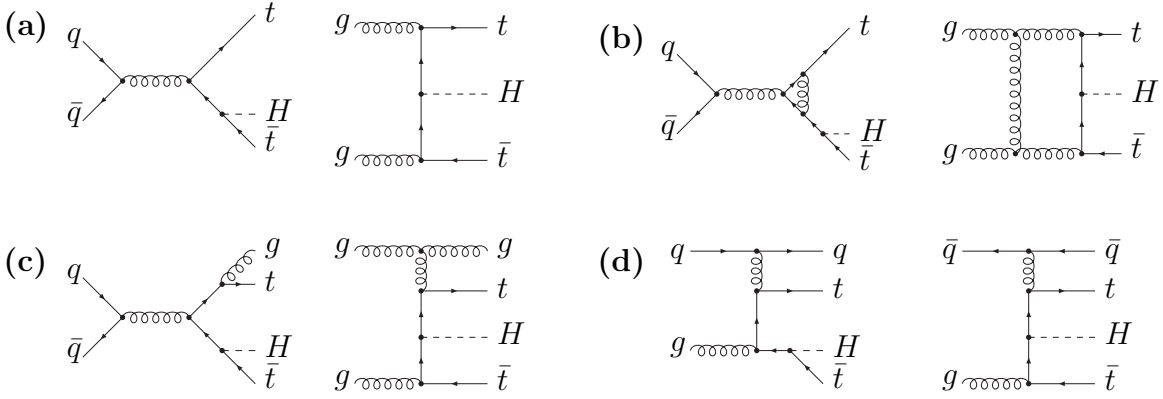
\begin{figure}
\SetScale{0.5}
\noindent
{\unitlength 0.5pt 
\begin{picture}(170,120)(-50,-10)
\ArrowLine(20, 80)(50,50)
\ArrowLine(50, 50)(20,20)
\Gluon(50,50)(100,50){4}{5}
\Vertex(50,50){2}
\Vertex(100,50){2}
\Vertex(125,25){2}
\DashLine(125, 25)(150,25){5}
\ArrowLine(100,50)(150,100)
\ArrowLine(125,25)(100, 50)
\ArrowLine(150, 0)(125, 25)
\put( 8,80){$q$}
\put( 8,20){$\bar q$}
\put(158, 95){$t$}
\put(158,20){$H$}
\put(158,-3){$\bar{t}$}
\put(-40, 85){{\bf (a)}}
\end{picture}
\hspace*{0em}
\begin{picture}(170,120)(-40,0)
\Gluon(50,100)(100,100){4}{5}
\Gluon(50,  0)(100,  0){4}{5}
\Vertex(100,100){2}
\Vertex(100,  0){2}
\Vertex(100, 50){2}
\DashLine(100, 50)(150,50){5}
\ArrowLine(100,100)(150,100)
\ArrowLine(100, 50)(100,100)
\ArrowLine(100,  0)(100, 50)
\ArrowLine(150,  0)(100,  0)
\put(158, 95){$t$}
\put(158,44){$H$}
\put(158,-5){$\bar{t}$}
\put(35,100){$g$}
\put(35,  0){$g$}
\end{picture}
}
\hspace*{3em}
{\unitlength 0.5pt 
\begin{picture}(170,120)(-50,0)
\ArrowLine(20, 80)(50,50)
\ArrowLine(50, 50)(20,20)
\Gluon(50,50)(100,50){4}{5}
\Vertex(50,50){2}
\Vertex(100,50){2}
\Vertex(130,20){2}
\Vertex(115,65){2}
\Vertex(115,35){2}
\DashLine(130, 20)(150,20){5}
\ArrowLine(115,65)(150,100)
\ArrowLine(100,50)(115, 65)
\ArrowLine(115,35)(100, 50)
\ArrowLine(130,20)(115,35)
\ArrowLine(150, 0)(130, 20)
\Gluon(115,65)(115,35){4}{3}
\put( 8,80){$q$}
\put( 8,20){$\bar q$}
\put(158, 95){$t$}
\put(158,15){$H$}
\put(158,-7){$\bar{t}$}
\put(-40, 85){{\bf (b)}}
\end{picture}
\hspace*{0em}
\begin{picture}(210,120)(-50,0)
\Gluon(50, 95)( 90, 95){4}{4}
\Gluon(50,  5)( 90,  5){4}{4}
\Gluon(90, 95)(140, 95){4}{5}
\Gluon(90,  5)(140,  5){4}{5}
\Gluon(90, 95)( 90,  5){4}{9}
\Vertex( 90, 95){2}
\Vertex( 90,  5){2}
\Vertex(140,  5){2}
\Vertex(140, 50){2}
\Vertex(140, 95){2}
\DashLine(140,50)(170,50){5}
\ArrowLine(140, 95)(170, 95)
\ArrowLine(140, 50)(140, 95)
\ArrowLine(140,  5)(140, 50)
\ArrowLine(170,  5)(140,  5)
\put(178, 90){$t$}
\put(178,44){$H$}
\put(178, 2){$\bar{t}$}
\put(35, 95){$g$}
\put(35,   2){$g$}
\end{picture}
} 
\\[2em]
{\unitlength 0.5pt 
\begin{picture}(170,120)(-50,0)
\ArrowLine(20, 80)(50,50)
\ArrowLine(50, 50)(20,20)
\Gluon(50,50)(100,50){4}{5}
\Vertex(50,50){2}
\Vertex(100,50){2}
\Vertex(125,25){2}
\Vertex(125,75){2}
\DashLine(125, 25)(150,25){5}
\ArrowLine(125,75)(150, 75)
\ArrowLine(100,50)(125, 75)
\ArrowLine(125,25)(100, 50)
\ArrowLine(150, 0)(125, 25)
\Gluon(125,75)(150,100){4}{3}
\put(158,100){$g$}
\put( 8,80){$q$}
\put( 8,20){$\bar q$}
\put(158, 70){$t$}
\put(158,20){$H$}
\put(158,-3){$\bar{t}$}
\put(-40, 85){\bf (c)}
\end{picture}
\hspace*{0em}
\begin{picture}(170,120)(-40,0)
\Gluon(50,100)(100,100){4}{5}
\Gluon(100,100)(150,100){4}{5}
\Gluon(100,100)(100,70){4}{3}
\Gluon(50,  0)(100,  0){4}{5}
\Vertex(100,100){2}
\Vertex(100, 70){2}
\Vertex(100,  0){2}
\Vertex(100, 35){2}
\DashLine(100, 35)(150,35){5}
\ArrowLine(100, 70)(150, 70)
\ArrowLine(100, 35)(100, 70)
\ArrowLine(100,  0)(100, 35)
\ArrowLine(150,  0)(100,  0)
\put(158,100){$g$}
\put(158, 65){$t$}
\put(158,30){$H$}
\put(158,-5){$\bar{t}$}
\put(35,100){$g$}
\put(35,  0){$g$}
\end{picture}
}
\hspace*{3em}
{\unitlength 0.5pt 
\begin{picture}(170,120)(-30,0)
\ArrowLine(50,100)(100,100)
\ArrowLine(100,100)(150,100)
\Gluon(100,100)(100,70){4}{3}
\Gluon(50, 25)(100, 25){4}{5}
\Vertex(100,100){2}
\Vertex(100, 70){2}
\Vertex(100, 25){2}
\Vertex(125, 25){2}
\DashLine(125, 25)(150, 25){5}
\ArrowLine(100, 70)(150, 70)
\ArrowLine(100, 25)(100, 70)
\ArrowLine(125, 25)(100, 25)
\ArrowLine(150, 0)(125, 25)
\put(158, 95){$q$}
\put(158, 65){$t$}
\put(158,20){$H$}
\put(158,-3){$\bar{t}$}
\put(35, 95){$q$}
\put(35, 25){$g$}
\put(-20, 85){\bf (d)}
\end{picture}
\hspace*{0em}
\begin{picture}(210,120)(-50,0)
\ArrowLine(100,100)(50,100)
\ArrowLine(150,100)(100,100)
\Gluon(100,100)(100,70){4}{3}
\Gluon(50,  0)(100,  0){4}{5}
\Vertex(100,100){2}
\Vertex(100, 70){2}
\Vertex(100,  0){2}
\Vertex(100, 35){2}
\DashLine(100, 35)(150,35){5}
\ArrowLine(100, 70)(150, 70)
\ArrowLine(100, 35)(100, 70)
\ArrowLine(100,  0)(100, 35)
\ArrowLine(150,  0)(100,  0)
\put(158, 65){$t$}
\put(158,30){$H$}
\put(158,-5){$\bar{t}$}
\put(35,  0){$g$}
\put(158, 95){$\bar q$}
\put(35, 95){$\bar q$}
\end{picture}
} 
\vspace*{0em}
\caption{A generic set of diagrams (a) for the Born level,
(b) one-loop corrections, (c) gluon radiation and (d) parton
splitting in the subprocesses $q\bar q, gg\to t\bar tH$ etc.}
\label{fig:diags}
\end{figure}

A survey of typical Feynman diagrams contributing to the processes
$p\bar p/pp\to\Pt\bar\Pt\PH+X$ in LO and NLO is shown in \reffi{fig:diags}.
In LO the hadron collisions proceed via $q\bar q$ annihilation
and $gg$ scattering at the parton level, as illustrated in 
\reffi{fig:diags}(a). In NLO, virtual corrections are induced by
one-loop diagrams, such as depicted in \reffi{fig:diags}(b), and
real corrections are induced by gluon bremsstrahlung and by
parton splittings $q\to qg$, $\bar q\to\bar qg$ in the
initial state, as shown in \reffis{fig:diags}(c) and (d), respectively.
Among the one-loop diagrams, 
the pentagons, which have five internal propagators in the loop,
are most complicated.
The idea of the analytical calculation and the strategy for a
numerically stable evaluation of these pentagons are sketched
in the next subsection. The most complicated part in the real corrections
concerns the extraction of the infrared (IR) singularities and
their proper cancellation; this is briefly described at the end of this
section.

\paragraph{Virtual corrections and pentagon diagrams}

The main complications in the calculation of the pentagon diagrams are
IR (soft and collinear) singularities. The most
convenient way to regularize these divergences is dimensional 
regularization.
As known for a long
time \cite{Melrose:1965kb}, in $D=4$ dimensions all 5-point functions
can be expressed in terms of 4-point functions, simplifying the
calculation considerably. 
In order to derive a smiliar reduction of 5-point functions to related
4-point functions in $D$ dimensions, first
the dimensionally regularized integral $E^{(D)}$ is translated into
another regularization scheme that is defined for $D=4$.
For instance, it is possible to endow all massless propagators in the loop with
an infinitesimal mass $\la$, defining the new integral
$E^{(\mathrm{mass},D)}$, which is identical to $E^{(D)}$ if $\la=0$.
In the next step, a related {\it well-defined} integral, denoted
$E^{(\mathrm{mass},D)}_{\mathrm{sing}}$, is determined that
possesses the same IR singularity structure as $E^{(\mathrm{mass},D)}$. This
integral is obtained by decomposing the integrand of the 5-point
function in the collinear/soft limit in terms of 3-point integrands
with regularization-scheme-independent kinematical prefactors.
The explicit construction of $E^{(\mathrm{mass},D)}_{\mathrm{sing}}$ is
described in \citere{pptth_long} in detail and illustrated in
\reffi{fig:Esing} for a specific example diagrammatically.
\bfi
\centerline{
\SetScale{.5}
\unitlength 1pt 
\begin{picture}(440,120)(0,0)
\put(80,92){$\left.\phantom{\rule{.1cm}{1.0cm}}\right|_{\,\mathrm{sing}}$}
\put(-5,70){\unitlength 1pt 
\begin{picture}(85,50)(0,0)
\Gluon(15, 95)(65,90){4}{4}
\Gluon(65, 10)(15, 5){4}{4}
\Gluon(65, 90)(115,90){4}{4}
\Gluon(115, 10)(65,10){4}{4}
\Gluon(65,10)(65,90){4}{6}
\Vertex(65,90){2}
\Vertex(65,10){2}
\ArrowLine(115, 10)(125,50)
\ArrowLine(125, 50)(115,90)
\DashLine(125, 50)(165,50){5}
\ArrowLine(165,  5)(115, 10)
\ArrowLine(115,90)(165,95)
\Vertex(115,90){2}
\Vertex(115,10){2}
\Vertex(125,50){2}
\end{picture}
}
\put(110,92){=}
\put(130,92){${f_1}\,\times$}
\put(135,70){\unitlength 1pt 
\begin{picture}(85,50)(0,0)
\Gluon(15, 95)(65,90){4}{4}
\Gluon(65, 10)(15, 5){4}{4}
\Gluon(65, 90)(125,50){4}{5}
\Gluon(125, 50)(65,10){4}{5}
\Gluon(65,10)(65,90){4}{6}
\Vertex(65,90){2}
\Vertex(65,10){2}
\DashLine(125, 50)(165,50){5}
\ArrowLine(165,  5)(125, 50)
\ArrowLine(125,50)(165,95)
\GCirc(125, 50){10}{0}
\end{picture}
}
\put(230,92){+}
\put(250,92){${f_2}\,\times$}
\put(250,70){\unitlength 1pt 
\begin{picture}(85,50)(0,0)
\Gluon(15, 95)(65,90){4}{4}
\Gluon(90, 30)(15, 5){4}{6}
\Gluon(65, 90)(115,90){4}{4}
\Gluon(90,30)(65,90){4}{5}
\Vertex(65,90){2}
\ArrowLine(90,30)(115, 90)
\DashLine(90,30)(165,50){5}
\ArrowLine(165,  5)(90, 30)
\ArrowLine(115,90)(165,95)
\Vertex(115,90){2}
\GCirc(90,30){10}{0}
\end{picture}
}
\put(345,92){+}
\put(365,92){${f_3}\,\times$}
\put(365,70){\unitlength 1pt 
\begin{picture}(85,50)(0,0)
\Gluon(15, 95)(90,70){4}{6}
\Gluon(65, 10)(15, 5){4}{4}
\Gluon(115, 10)(65,10){4}{4}
\Gluon(65,10)(90,70){4}{5}
\Vertex(65,10){2}
\ArrowLine(115, 10)(90,70)
\DashLine(90,70)(165,50){5}
\ArrowLine(165,  5)(115, 10)
\ArrowLine(90,70)(165,95)
\Vertex(115,10){2}
\GCirc(90,70){10}{0}
\end{picture}
}
\put(135,18){+}
\put(155,18){${f_4}\,\times$}
\put(170,-5){\unitlength 1pt 
\begin{picture}(85,50)(0,0)
\Gluon(15, 95)(65,90){4}{4}
\Gluon(90, 30)(15, 5){4}{6}
\Gluon(65, 90)(115,90){4}{4}
\Gluon(90,30)(65,90){4}{5}
\Vertex(65,90){2}
\ArrowLine(90,30)(115, 90)
\DashLine(115,90)(165,50){5}
\ArrowLine(165,  5)(90, 30)
\ArrowLine(115,90)(165,95)
\GCirc(115,90){10}{0}
\GCirc(90,30){10}{0}
\end{picture}
}
\put(270,18){+}
\put(290,18){${f_5}\,\times$}
\put(300,-5){\unitlength 1pt 
\begin{picture}(85,50)(0,0)
\Gluon(15, 95)(90,70){4}{6}
\Gluon(65, 10)(15, 5){4}{4}
\Gluon(115, 10)(65,10){4}{4}
\Gluon(65,10)(90,70){4}{5}
\Vertex(65,10){2}
\ArrowLine(115, 10)(90,70)
\DashLine(115,10)(165,50){5}
\ArrowLine(165,  5)(115, 10)
\ArrowLine(90,70)(165,95)
\GCirc(115,10){10}{0}
\GCirc(90,70){10}{0}
\end{picture}
}
\end{picture}
\SetScale{1}
} 
\caption{Diagrammatical decomposition of the singular part of 
a pentagon diagram in terms of triangle diagrams, where the blobs
result from shrinking propagators to a point and the $f_i$ denote
simple kinematical prefactors.}
\label{fig:Esing}
\efi
The difference of the two integrals,
$E^{(\mathrm{mass},D)}-E^{(\mathrm{mass},D)}_{\mathrm{sing}}$, has a
uniquely-determined integrand and it is 
regularization-scheme independent, \ie the limits $D\to 4$ and $\la\to 0$ 
commute in this quantity. In total, we have generated in this way
the relation
\beq
E^{(D)}-E^{(D)}_{\mathrm{sing}} + \O(D-4) =
E^{(\mathrm{mass},D=4)}-E^{(\mathrm{mass},D=4)}_{\mathrm{sing}} +\O(\la),
\label{eq:Etrans}
\eeq
where $\lambda=0$ on the l.h.s.\ and $D=4$ on the r.h.s.
Consequently, solving Eq.~\refeq{eq:Etrans} for $E^{(D)}$,
\beq
E^{(D)} = E^{(D)}_{\mathrm{sing}} + \left[
E^{(\mathrm{mass},D=4)}-E^{(\mathrm{mass},D=4)}_{\mathrm{sing}} 
\right]+\dots,
\label{eq:Edim}
\eeq
this integral is expressed (up to irrelevant terms indicated by the
ellipsis) in terms of four-dimensional integrals and $D$-dimensional
3-point functions.

The translation of $D$-dimensional integrals into four dimensions
works not only for complete Feynman diagrams, but 
also for individual scalar and
tensor integrals. As shown in \citere{Melrose:1965kb}, for $D=4$
the scalar 5-point function $E_0$ can be expressed in terms of a linear
combination of five 4-point functions. Thus, Eq.~\refeq{eq:Edim}
expresses $E_0^{(D)}$ in terms of scalar 3- and 4-point functions.
For the evaluation of 5-point tensor integrals, two entirely different
methods have been used. In one calculation the well-known
Passarino--Veltman algorithm \cite{Passarino:1979jh}
is adopted to algebraically reduce the
tensor coefficients recursively to tensors of lower rank, eventually
leading to scalar integrals. This procedure requires to solve a set
of linear equations for each tensor rank and, in this way, adds a 
factor of an inverse Gram determinant in each step. 
At the phase-space boundary the Gram determinants vanish, since
the momenta that span the tensors become linearly dependent.
Near the phase-space boundary this leads to numerical instabilities
that are controlled by careful extrapolation out
of the safe inner phase-space domains. 
The second calculation, however, 
avoids the appearance of leading Gram determinants
(i.e.\ Gram deteminants formed by four momenta) by using the method
of \citere{Etensor}, which is based on a generalization of the strategy
\cite{Melrose:1965kb} for scalar 5-point functions and reduces 5-point
tensor coefficients to 4-point integrals directly. Applying this alternative
renders the virtual correction near the phase-space boundary much more
stable than in the usual Passarino--Veltman approach, and the 
extrapolation from the inner part of phase space turns out to be
practically unnecessary. The results obtained by the two methods
mutually agree with each other.

\paragraph{Real corrections and IR singularities}

Although the $2\to 4$ scattering matrix element of the real NLO
correction $\sigma^{\mathrm{real}}$ [see \reffis{fig:diags}(c) and (d)]
are quite involved, they still could be calculated by conventional
trace techniques or by using the {\sc Madgraph} package
\cite{Stelzer:1994ta}. The IR (soft and collinear) singularities
appearing in the phase-space integral of the squared matrix elements
have to be treated in dimensional regularization and they must be
extracted before
the numerical integration over the four-particle phase space.
To this end, a generalization \cite{Catani:2002hc}
of the dipole subtraction
formalism \cite{Catani:1996jh} to massive quarks has been adopted.
In this formalism
the singularities of the cross section $\sigma^{\mathrm{real}}$ are mapped
onto a suitably chosen auxiliary cross section $\sigma^{\mathrm{sub}}$
so that the difference
$\sigma^{\mathrm{real}}-\sigma^{\mathrm{sub}}$ can safely be
integrated numerically in four dimensions.  
Moreover, $\sigma^{\mathrm{sub}}$
is still simple enough to allow for an analytical integration over
the singular regions in phase space. The result of this integration
consists of LO cross sections dressed by universal functions that
contain the singularities.
This integrated cross section is decomposed into a part
$\bar\sigma^{\mathrm{sub}}_1$ that, defined on configurations with LO
kinematics, cancels the soft and collinear singularities of the
virtual corrections; and a second part $\bar\sigma^{\mathrm{sub}}_2$
that includes the singularities from initial-state parton splitting,
which are absorbed in the renormalization of the parton
densities. Thus the total NLO correction $\Delta\sigma^{\mathrm{NLO}}$
can be written as the sum
\beq
\Delta\sigma^{\mathrm{NLO}}  = 
\left[\sigma^{\mathrm{real}}-\sigma^{\mathrm{sub}}\right]
+ \left[\sigma^{\mathrm{virtual}}+\bar\sigma^{\mathrm{sub}}_1\right]
+ \left[\sigma^{\mathrm{part}}+\bar\sigma^{\mathrm{sub}}_2\right],
\eeq
in which each bracket is separately finite.

\section{Some numerical results}

\begin{figure}
\centerline{
\setlength{\unitlength}{1cm}
\begin{picture}(14.0,9.5)
\put(-0.5,7.1){{\bf (a)}}
\put(9.3,5.9){\small MRST PDFs}
\put(-0.7,-5.7){\includegraphics{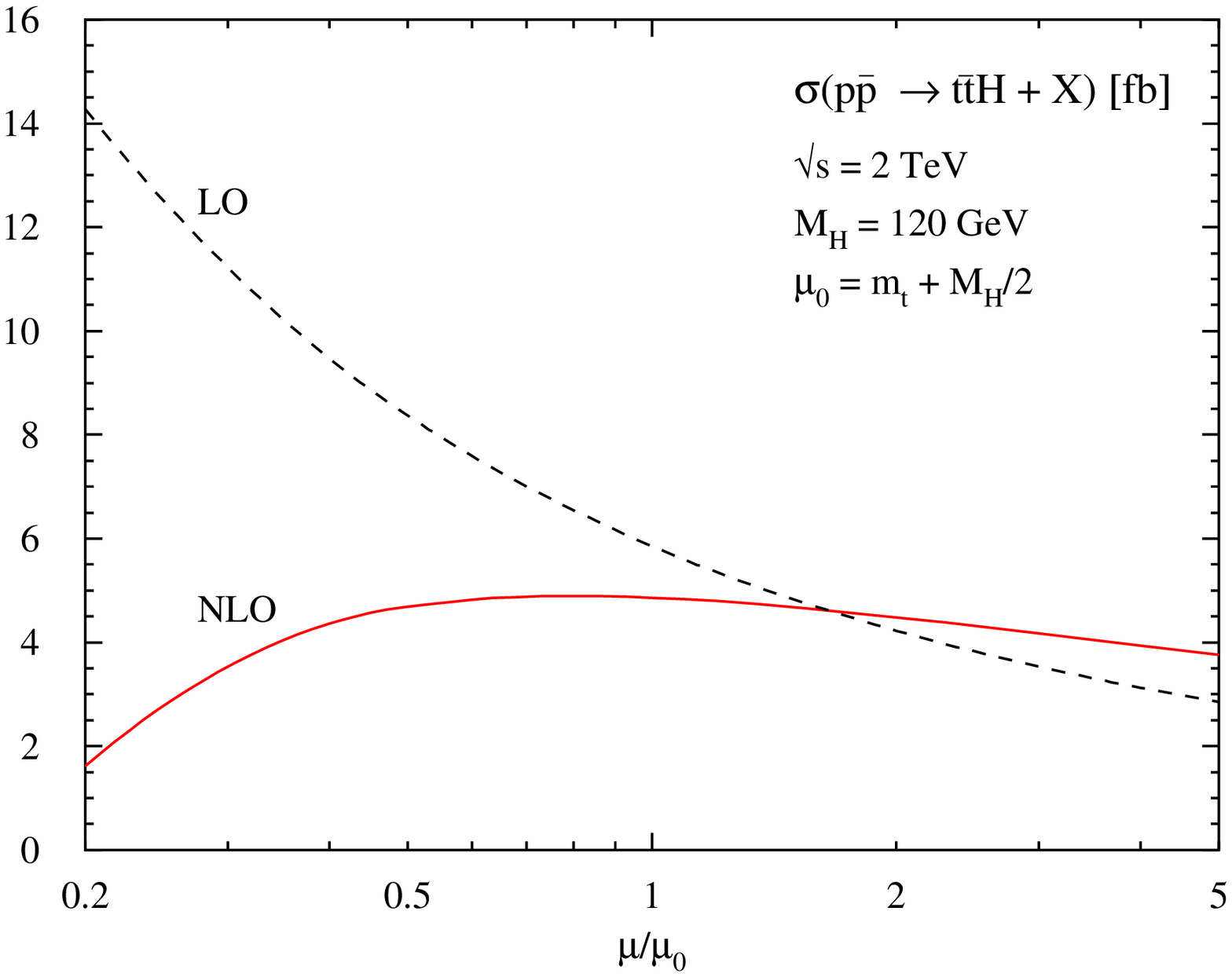}}
\end{picture}
} 
\vspace*{1em}
\centerline{
\setlength{\unitlength}{1cm}
\begin{picture}(14.0,9.5)
\put(-0.5,7.1){{\bf (b)}}
\put(9.3,5.9){\small MRST PDFs}
\put(-0.7,-5.7){\includegraphics{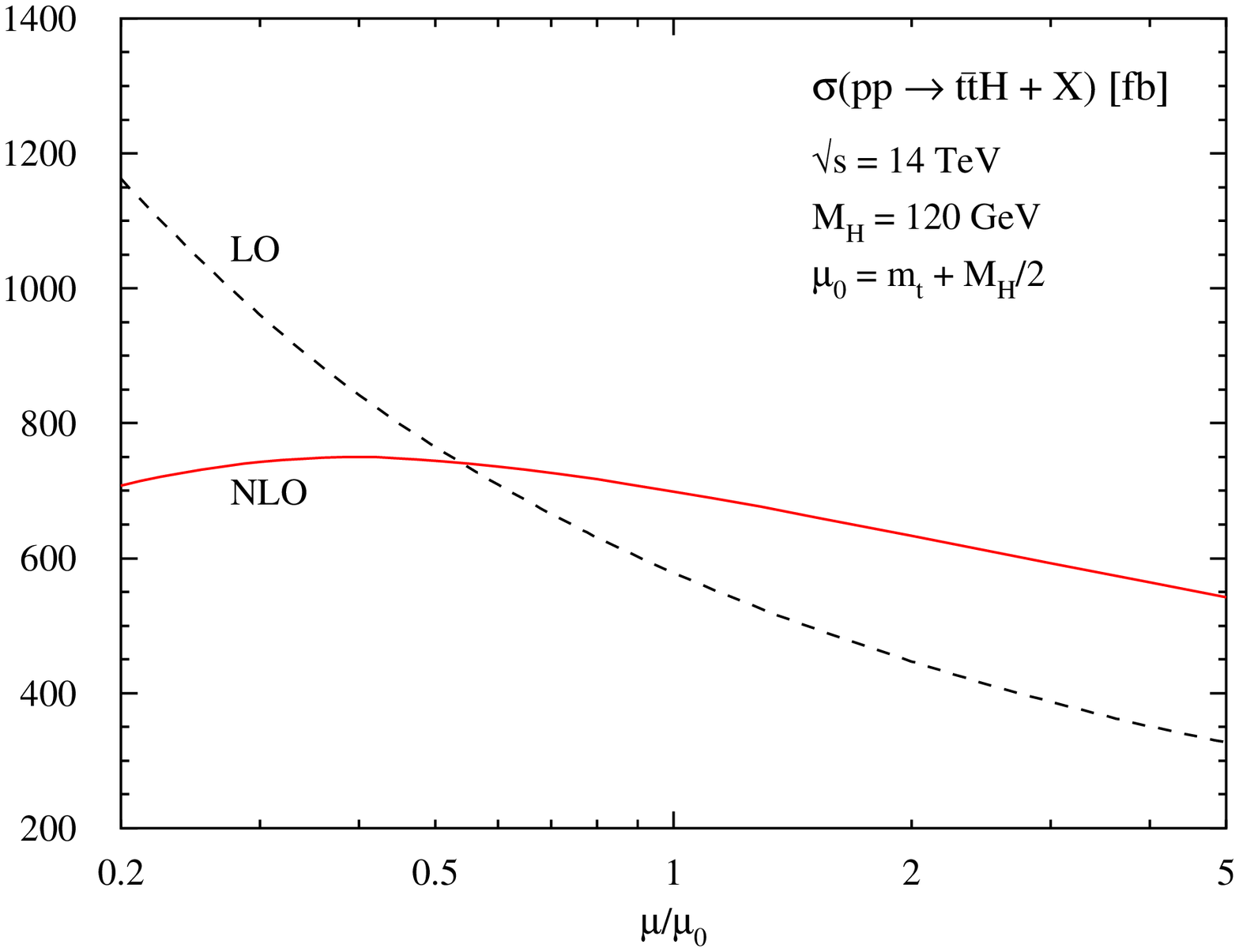}}
\end{picture}
} 
\caption{Variation of the LO and NLO cross sections with the renormalization
         and factorization scales for 
	(a) $\,p\bar p\to t\bar tH+X$ at Tevatron and 
	(b) $\,pp\to t\bar tH+X$ at the LHC.}
\label{fig:cs_mu}
\end{figure}

Figure~\ref{fig:cs_mu} shows the total cross sections for $\Pt\bar\Pt\PH$
production at the Tevatron and the LHC for a fixed Higgs-boson mass
of $\MH=120\GeV$ as function of the renormalization and factorization
scales, which have been taken at the common value $\mu$. 
\begin{figure}
\setlength{\unitlength}{1cm}
\centerline{
\begin{picture}(14,10.3)
\put(-0.5,7.1){{\bf (a)}}
\put(-0.7,-5.3){\includegraphics{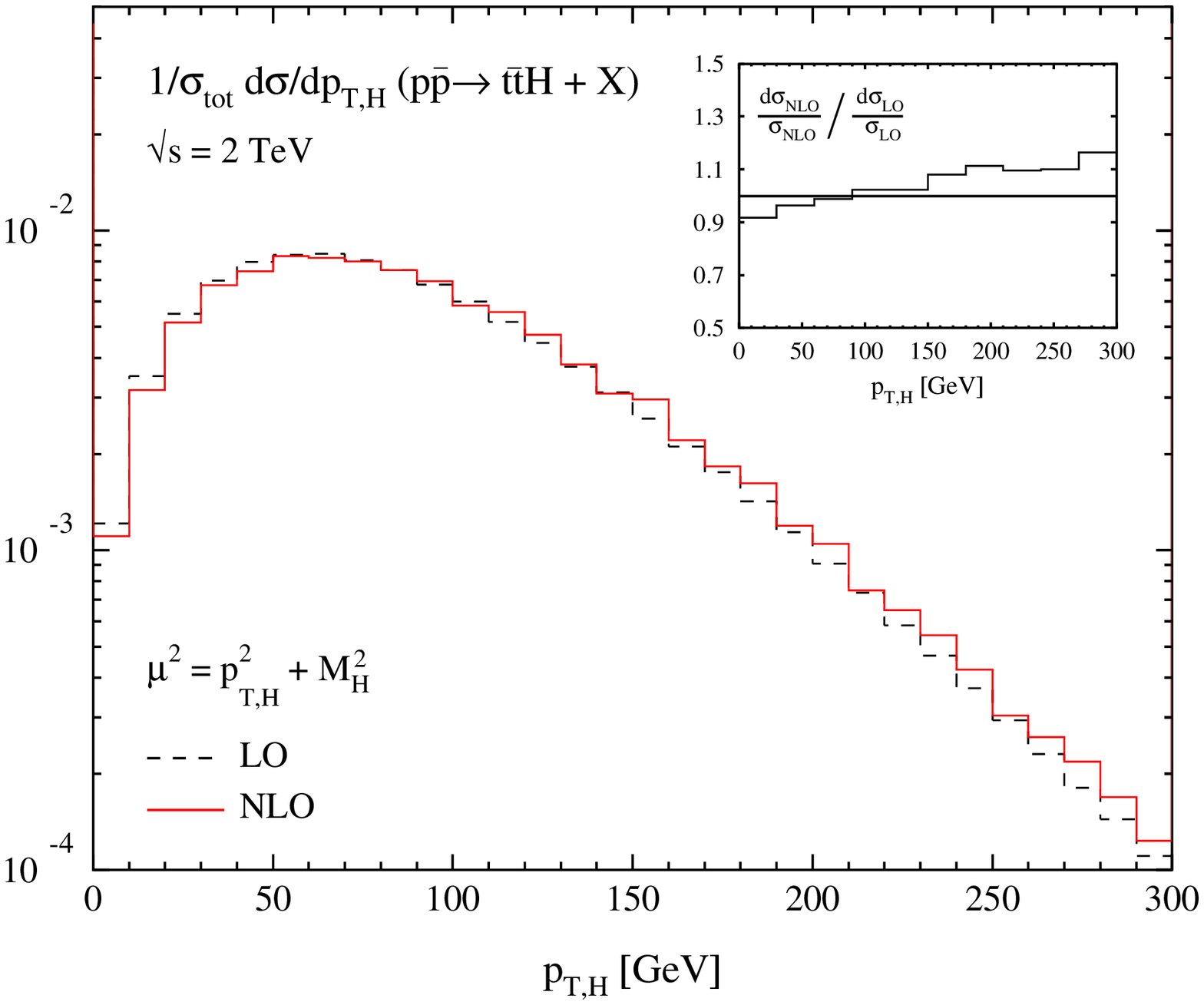}}
\end{picture}
} 
\vspace*{1em}
\centerline{
\begin{picture}(14,10.3)
\put(-0.5,7.1){{\bf (b)}}
\put(-0.7,-5.4){\includegraphics{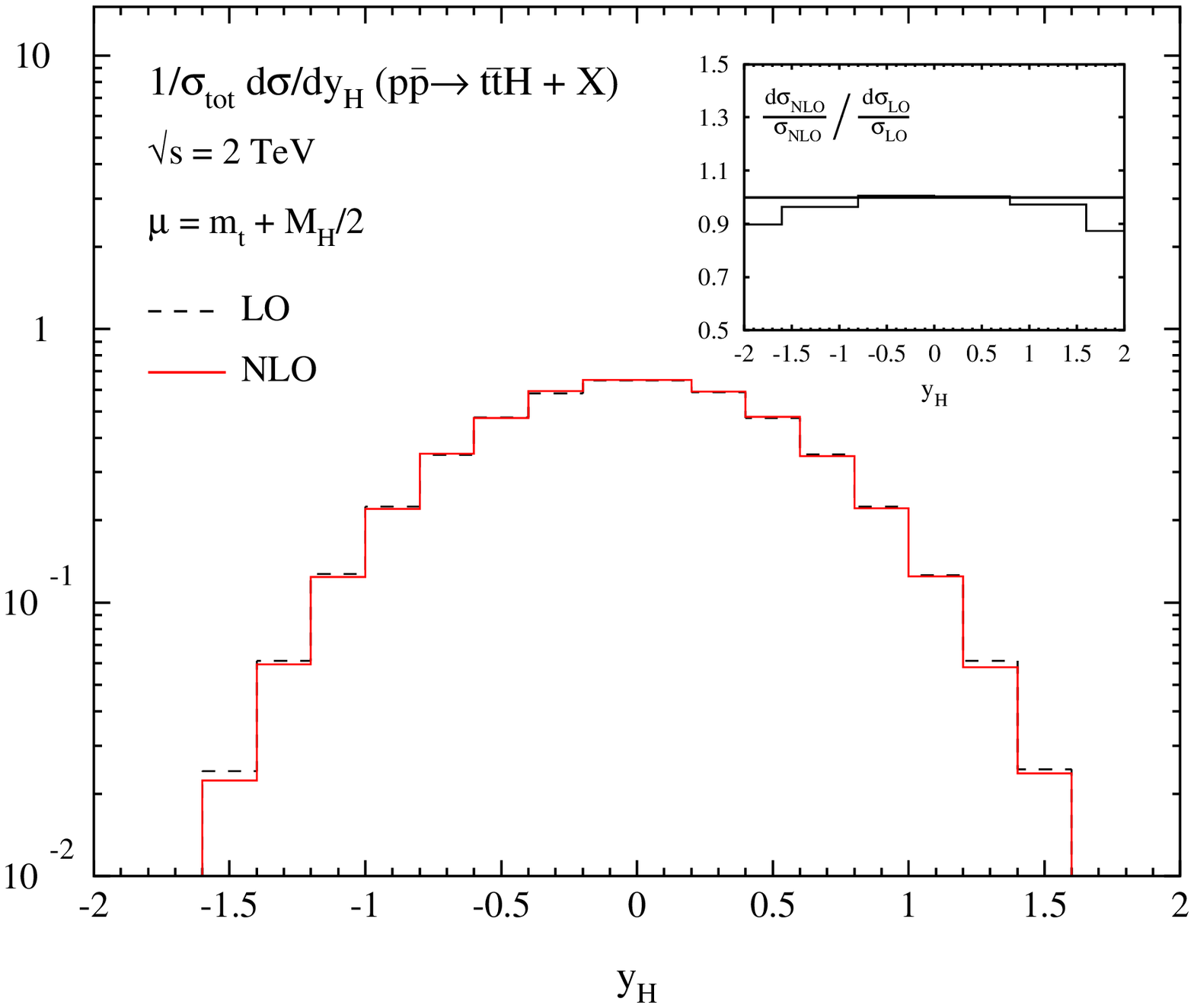}}
\end{picture}
} 
\caption{(a) Normalized transverse-momentum distribution and
	(b) rapidity distribution of the Higgs boson in LO 
         and NLO for $\,p\bar p\to t\bar tH+X$ at the Tevatron ($\MH=120\GeV$).}
\label{fig:pTH_Tev}
\end{figure}
It has been checked that no accidental compensations
of scale dependences are introduced by this identification.
In the vicinity of the central
scale $\mu_0=(2m_t+M_H)/2$, the NLO cross sections are remarkably
stable with very little variation for $\mu$ between $\sim\mu_0/3$ and
\mbox{$\sim 3\mu_0$}, in contrast to the LO approximation for which the
production cross section changes by more than a factor 2 within the
same interval. The relative corrections, which are
quantified by $K$ factors, $K=\sigma_{\NLO}/\sigma_{\LO}$
(with all quantities calculated consistently in NLO and LO, respectively)
are nearly constant in the Higgs-boson mass range $\MH=100$--$250\GeV$.

Although at the Tevatron the cross section is strongly
dominated by $q\bar q$ annihilation for scales
$\mu\sim\mu_0$, the proper study of the scale dependence requires
inclusion of the $gg$, $gq$, and $g\bar q$ channels. If $\mu$ is
chosen too low, large logarithmic corrections spoil the convergence of
perturbation theory, as indicated by the negative NLO cross section
for $\mu \lsim \mu_0/5$.  
Apparently the $K$ factor varies from $\sim 0.8$ at the
central scale $\mu=\mu_0$ to $\sim 1.0$ at the threshold scale
$\mu=2\mu_0$. As explained in \citere{Beenakker:2001rj},
the value of the $K$ factor below unity can be understood intuitively in
the fragmentation picture proposed in Ref.\cite{Dawson:1998im},
although the kinematical limit justifying this
approximation is not yet realized at the Tevatron.

At the LHC the major part of the cross section is due to the
$gg$ channel, which gives rise to increased gluon radiative corrections.
For the central scale $\mu_0$ the relative corrections amount to 
$K\sim 1.2$, increasing to $\sim 1.4$ at the 
threshold value $\mu=2\mu_0$. Again the fragmentation picture 
\cite{Dawson:1998im} is approximately compatible with the complete
NLO prediction.

Finally, as illustrating examples for final-state distributions,
\reffi{fig:pTH_Tev} shows the trans\-verse-mo\-men\-tum 
and rapidity distributions 
of the Higgs boson for $\,p\bar p\to t\bar tH+X$ at the Tevatron in LO and
NLO approximation. 
Note that for the distribution in the transverse Higgs-boson momentum
$p_{{\mathrm{T}},\PH}$, the scale $\mu$ is not taken constant but 
defined by $\mu^2=p_{{\mathrm{T}},\PH}^2+\MH^2$; this choice is more natural
for large transverse momenta. 
The ratio of the (normalized) NLO and LO
distributions reveals that the mere rescaling of the LO distribution
by a constant $K$ factor would reproduce the NLO distribution only
within $\pm 10\%$ in the most important ranges.
For the $p_{{\mathrm{T}},\PH}$ distribution
the increase of the ratio $\rd\sigma_{\NLO}/\rd\sigma_{\LO}$ with
increasing $p_{{\mathrm{T}},\PH}$ can be traced back to the scale variations
of $\rd\sigma_{\NLO}$ and $\rd\sigma_{\LO}$, since $\mu$ rises with
increasing $p_{{\mathrm{T}},\PH}$ and $\rd\sigma_{\LO}$ decreases with
increasing $\mu$ faster than $\rd\sigma_{\NLO}$. 
The respective distributions
at the LHC, which are displayed in \citere{pptth_long},
show the same qualitative behaviour, but in the $p_{{\mathrm{T}},\PH}$
distribution
the ratio $\rd\sigma_{\NLO}/\rd\sigma_{\LO}$ varies even more strongly with
$p_{{\mathrm{T}},\PH}$.

\section{Conclusion}

The strong scale dependence of the LO cross sections for
$p\bar p/pp\to t\bar tH+X$ is drastically reduced by the NLO calculation,
and the theoretical predictions are stabilized. Thus,
the NLO cross sections can serve as a solid base for experimental
analyses at the Tevatron and the LHC. This improvement is also observed 
for the final-state Higgs transverse-momentum and rapidity distributions.
\vspace{2em}


\end{document}